# Holographic EUV Lithography at 40 nm Resolution

Ziqi Li[1,2,3,4], Iason Giannopoulos[2], Lisong Dong[1,3,4], Dimitrios Kazazis[2], Xu Ma[5], Zongqiang Yu[6], Zhiyuan Niu[6], Yasin Ekinci[2],Yayi Wei[1,3,4,*] and Iacopo Mochi[2,*]

[1] *EDA Center, Institute of Microelectronics of Chinese Academy of Sciences, Beijing 100029, China*

[2] *Laboratory for X-ray Nanoscience and Technologies, Paul Scherrer Institute, Villigen 5232, Switzerland*

[3] *State Key Laboratory of Fabrication Technologies for Integrated Circuits, Beijing 100029, China*

[4] *University of Chinese Academy of Sciences, Beijing 100049, China*

[5] *Key Laboratory of Photoelectronic Imaging Technology and System of Ministry of Education of China, School of Optics and Photonics, Beijing Institute of Technology, Beijing, 100081, China*

[6] *Hyper Optics, Beijing, 100176, China*

* iacopo.mochi@psi.ch, weiyayi@ime.ac.cn

**Abstract.**

Extreme ultraviolet (EUV) lithography is the cornerstone of the fabrication of advanced integrated circuits at the 7-nm node and beyond, but its reliance on multi-element reflective projection optics makes it inaccessible for small-scale research and prototyping. EUV interference lithography (EUV-IL) provides a lensless alternative but is intrinsically restricted to periodic structures. Here we demonstrate EUV holographic lithography (EUV-HL) as a lensless route to arbitrary, non-periodic, curvilinear patterning at the EUV wavelength of 13.5 nm. We introduce an inverse-design framework for computer-generated holograms that captures the dominant physical effects of EUV mask diffraction within a shift-invariant convolution model that is tractable for full mask layouts. Using this framework, we design and fabricate transmissive holographic masks by direct-write electron-beam lithography in hydrogen silsesquioxane, expose them with synchrotron-generated EUV radiation, and print target layouts with critical dimensions down to 40 nm, nearly an order of magnitude finer than the previous state of the art in EUV-HL. The demonstrated combination of sub-50 nm resolution, curvilinear design freedom, and a lensless optical setup establishes EUV-HL as a uniquely flexible tool for nanostructure prototyping at EUV wavelengths, and provides a natural pathway to non-periodic pattern prototyping at beyond-EUV (BEUV) wavelengths, which is currently inaccessible to interference-based methods.

# 1 Introduction

The rapid growth of frontier technologies, including artificial intelligence, autonomous driving, and high-performance computing, relies critically on continued advances in semiconductor manufacturing[1,2]. Scaling the density and performance of integrated circuits (ICs) into the 7-nm technology node and beyond demands lithographic tools capable of patterning ever-smaller features with high fidelity and throughput. Extreme ultraviolet (EUV) lithography has therefore become an indispensable technology for sustaining Moore's law and enabling next-generation devices[3,4].

Despite its success in high-volume manufacturing, modern EUV scanners remain extraordinarily complex and expensive systems[5,6]. Among the most challenging subsystems is the projection optics: a multi-element reflective lens assembly that must operate with minimal flare, aberration, and wavefront distortion at a wavelength of 13.5 nm[7,8]. Achieving such performance requires mirror substrates and multilayer coatings with surface figures controlled to sub-nanometer accuracy and surface roughness below 0.1 nm[9], pushing optical fabrication and metrology to their physical and economic limits. These stringent requirements, combined with the massive engineering effort needed for source power, contamination control, and precision mechanics, drive the escalating cost and limited scalability of EUV lithography[5,10].

In parallel with production-level EUV lithography, EUV interference lithography (EUV-IL) has emerged as a powerful technique for resist screening, process development, and rapid patterning of periodic structures[11–16]. By interfering two or more coherent EUV beams, EUV-IL enables the generation of high-contrast line-space or contact-hole patterns over large areas without the need for complex projection optics or device-specific masks. Its simplicity, coherence-based resolution, and ability to decouple exposure from mask design have made it an essential tool for photoresist characterization and for benchmarking EUV process windows[12,17]. However, EUV-IL is intrinsically limited to periodic structures, restricting its usefulness for prototyping device-relevant geometries.

To overcome this limitation, EUV holographic lithography (EUV-HL) offers a compelling extension of interference-based methods. In EUV-HL, a computer-generated hologram (CGH) mask is illuminated with coherent EUV light to synthesize a desired aerial image directly through diffraction. The holographic mask is explicitly designed such that its diffracted field reconstructs the target intensity distribution at the wafer plane, enabling arbitrary pattern formation without relying on projection optics. In principle, this approach retains the low-cost and high-coherence advantages of interference lithography while dramatically expanding the accessible design space.

Previous works demonstrate successful application of HL for the fabrication of periodical 3D structures[18–21], while the pursuit of imaging with higher resolution remains challenging. Prior attempts of HL have been limited under i-line[22] and visible light[23] regime, and efforts of EUV-HL was less-than-ideal[24,25]. Early demonstrations largely focused on simple or low-resolution holograms, constrained by insufficient mask modeling, limited coherence control, and fabrication challenges. So far, the best resolution for EUV holographic lithography was 372 nm obtained on elbow patterns with a pitch of 602 nm[26]. Unfortunately, this is below the required resolution for meaningful device prototyping, preventing broader adoption of holographic approaches.

Despite this promise, prior attempts at EUV-HL have been limited[24,25]. Early demonstrations largely focused on simple or low-resolution holograms, constrained by insufficient mask modeling, limited coherence control, and fabrication challenges. So far, the best resolution for EUV holographic lithography was 372 nm obtained on elbow patterns with a pitch of 602 nm[26]. Unfortunately, this is below the required resolution for meaningful device prototyping, preventing broader adoption of holographic approaches.

In this work, we close this gap and establish EUV-HL as a practical, lensless route to arbitrary nanopatterning at EUV wavelengths. The central contribution is an inverse-design framework that computes the holographic mask directly from the target layout while accurately modeling the interaction between partially coherent, polychromatic EUV illumination and a three-dimensional absorber stack. By formulating the mask three-dimensional (M3D) response as a shift-invariant convolution and combining it with angular-spectrum propagation, the forward model becomes tractable on full mask layouts, and the corresponding inverse problem can be solved by gradient-based optimization. Using masks designed with this framework and fabricated by direct-write electron-beam lithography in hydrogen silsesquioxane, we demonstrate, for the first time, EUV holographic lithography with critical dimensions down to 40 nm, on arbitrary, non-periodic, curvilinear layouts that lie outside the reach of EUV-IL. Beyond the order-of-magnitude resolution improvement, the results define a design space that is qualitatively distinct from both projection EUV and EUV-IL, and that maps naturally onto the kinds of devices like metasurfaces, photonic crystals, quantum emitter arrays, and superconducting nanowire structures, for which arbitrary EUV-scale patterning is currently a bottleneck.

## 2 Experimental Concept and Setup

EUV-HL eliminates the complexity and fabrication challenges associated with EUV projection optics by employing a single diffracting element: the holographic mask. Figure 1 contrasts the two approaches, showing the conventional EUV projection system (a) alongside the EUV-HL configuration (b). In EUV-HL, a plane wave illuminates a transmissive mask, and the diffracted field propagates in free space to the wafer plane, where its intensity distribution reproduces the target aerial image.

Designing such a mask, however, is non-trivial. Accurate modeling must account for the three-dimensional structure of the mask, the optical properties of both the membrane and the absorber, the spectral distribution of the illumination, and the constraints imposed by the fabrication process. Given a target pattern and the experimental parameters, we first compute the corresponding mask pattern, fabricate the mask according to the design, expose it to synchrotron-generated EUV radiation, and transfer the pattern to a photoresist-coated wafer. The imaging model, mask computation, and mask fabrication are discussed in detail in Section 4.

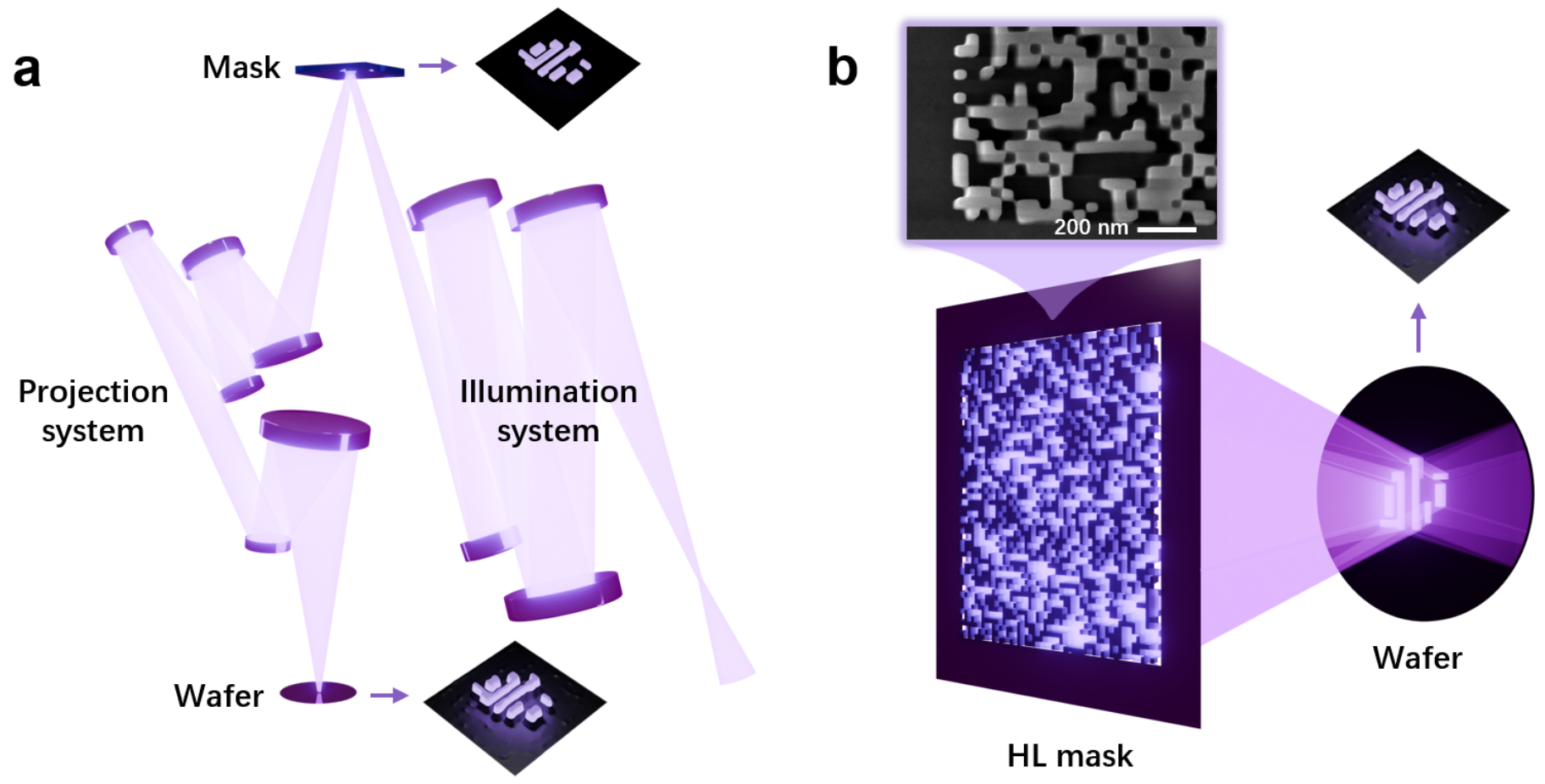

*Figure 1. Conceptual designs of: (a) the conventional EUV projection lithography system, and (b) the EUV-HL system. By utilizing diffraction-based imaging, EUV HL removes the need for EUV projection optics, remarkably simplifying the system without a considerable loss in resolution.*

### 2.1 The holographic mask

The mask consists of a silicon support frame holding an 80 nm-thick $Si_3N_4$ membrane, coated with a 200 nm-thick layer of hydrogen silsesquioxane (HSQ) that was patterned by electron-beam lithography to form the absorber. The SEM image in Figure 1(b) shows a representative portion of the resulting HSQ mask pattern. HSQ was chosen as the absorber for its compatibility with high-resolution e-beam patterning, its stability under EUV irradiation, and the fact that it can be used directly without an additional pattern-transfer step (see Section 4.3). The HSQ layer has a transmission of about 13.5% at 13.5 nm.

### 2.2 Exposure setup and imaging resist

The EUV-HL exposures were carried out at the multi-purpose beamline BL05U-A1 of the National Synchrotron Radiation Laboratory (NSRL) in Hefei, China. The beamline delivers a spatially coherent EUV beam with a spectral bandwidth of approximately 4% (FWHM). The end station, originally designed for interference lithography, enables precise positioning and alignment of the mask relative to the sample. It operates under cleanroom conditions and is mechanically and thermally isolated to suppress external vibrations and thermal drift, thereby maximizing the stability of the aerial image throughout the exposure.

To assess the performance and capabilities of EUV-HL, we used HSQ as the imaging resist on the wafer as well. HSQ is a high-resolution resist widely employed in electron-beam lithography that is also sensitive to EUV radiation[27]. Although its EUV sensitivity falls roughly an order of magnitude short of the requirements for industrial EUV lithography, resulting in prohibitively long exposure times for conventional EUV scanners, it serves as an excellent benchmarking resist for resolution studies. In our experiments, the combination of the synchrotron source flux and the HL mask geometry resulted in exposure times of 10–20 seconds.

The simulations we performed on elbow patterns show that the contrast of the holographic images is 40% at 80 nm pitch and 40 nm CD and increases to 50% for elbows with pitch 140 nm and 70 nm CD. The contrast of the aerial image affects the usable dose range of the lithography process. The steep dose to clear curve of HSQ, is useful to mitigate this effect which makes it an ideal resist for this study.

## 3 EUV Holographic Lithography Demonstration

The masks fabricated for this experiment feature a 2 × 2 mm² patterned membrane area, incorporating several holographic patterns designed to produce aerial images of test structures at a range of focal planes. The smallest design feature on the mask is 40 nm and the working distance is 200 μm.

Figure 2 presents representative SEM images collected on one of the exposed wafers, together with the corresponding regions of the mask that generated the respective aerial images. The results confirm that the holographic mask is capable of reproducing complex, non-periodic patterns, including curvilinear structures.

The first row of Figure 2 shows the mask pattern (a), the SEM image of the printed structure (b), and a magnified detail of the image (c). The printed pattern is a random Manhattan-geometry layout that mimics the logic layer of an integrated circuit, with a critical dimension (CD) of 70 nm.

The second row of Figure 2 shows the mask pattern and the printed image of an array of isolated glyphs with a CD of 80 nm. It is important to note that the mask pattern is generally larger than the printed image field of view. This allows the mask to encode high spatial frequency information that contributes to improving the resolution of the printed pattern. The ratio between the mask size and the image field of view is governed by the target resolution, the wavelength, and the working distance between the mask and the wafer, in a manner analogous to how the numerical aperture of a projection system determines the resolution of the aerial image.

Figure 2(g) shows the portion of the mask that generates the logo of the Institute of Microelectronics of the Chinese Academy of Sciences (IMECAS), while Figure 2(h) shows the corresponding SEM image of the logo printed on the wafer, and Figure 2(i) provides a magnified view of the character '科' (meaning 'science' in Chinese). This result further highlights EUV-HL's capability to reproduce curvilinear geometries, which broadens the design freedom available to IC designers and supports the future application of design-technology co-optimization (DTCO).

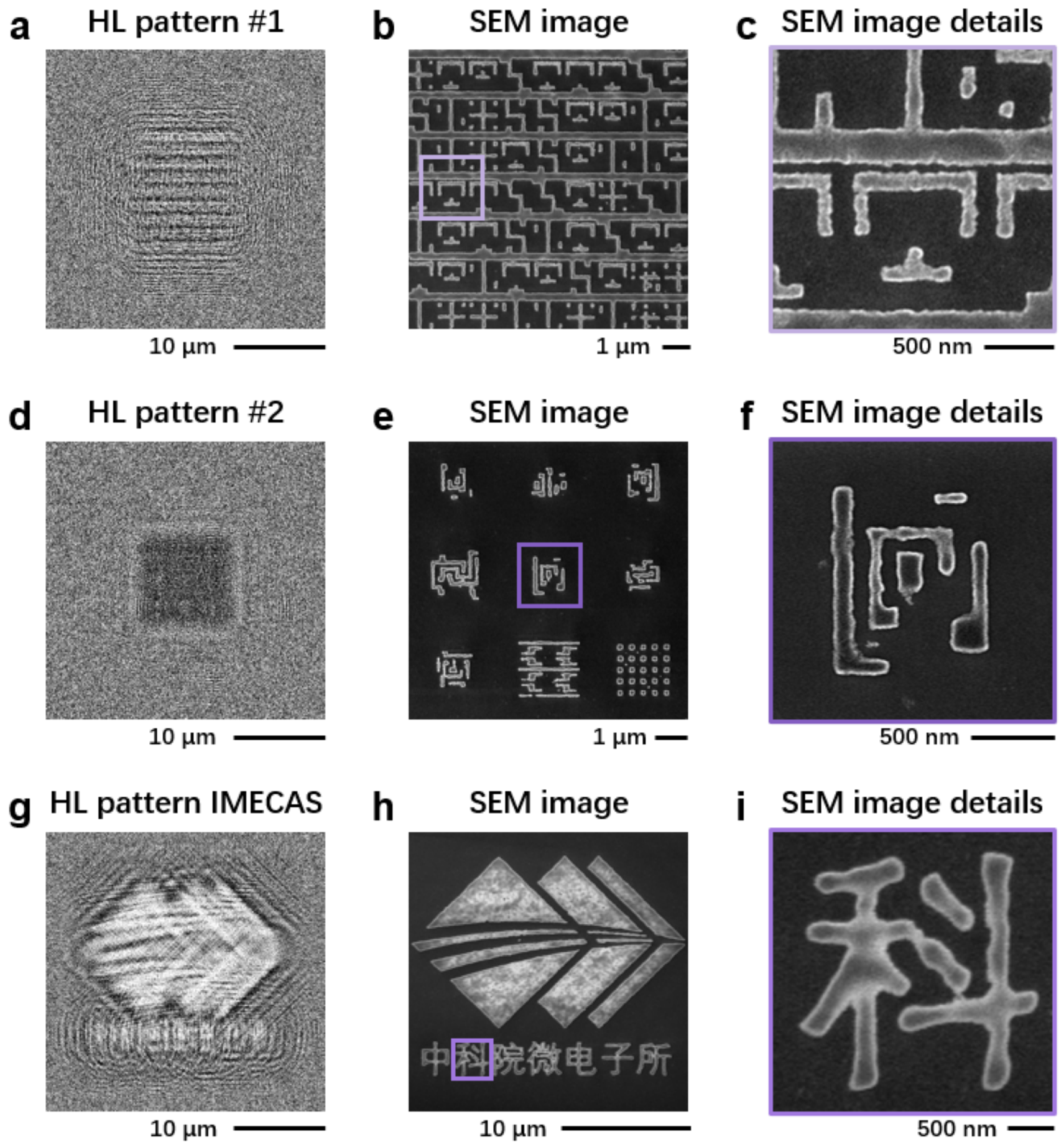

*Figure 2. The images in a, d and g show the pattern over three regions of the same mask while b, e and h are the corresponding SEM images of the developed wafer. c, f and i show a detail of the SEM images to give a qualitative reference about the printed pattern resolution.*

For a deeper look into the performance of the method, we superimposed the contour of the aerial image generated by the holographic mask as predicted by our simulation, to the corresponding SEM image of the printed wafer. While the aerial images are still just an approximation of the target layout, they match the printed pattern quite well.
To compare the SEM image of the printed pattern to the simulated aerial images we binarized them using a fixed threshold and we compared the mean square error of their difference. In this context, the mean square error (MSE) is defined as:

$$MSE = \frac{1}{N}\sum(T - R)^2,$$

where T is the array representing the binarized SEM image, R is the simulated resist pattern and N is the sum of T, to compare the simulation error of different cases with various pattern density.
The results for the image shown in Figure 3 are summarized in the following table:

*Table 1. Measured MSE for the patterns shown in Figure 3. The MSE was calculated on the difference between the binarized SEM images and the corresponding simulated resist profiles.*

| **Pattern** | **a** – Glyph | **b** – Tip to line | **c** – Glyph | **d** – Elbow |
|---|---|---|---|---|
| **CD** | 40 nm | 40 nm | 80 nm | 50 nm |
| **MSE** | 0.2282 | 0.2881 | 0.1513 | 0.3287 |

This demonstrates that the mask is performing as we expect.

In Figure 3, the first column shows four different target patterns with arrays of structures with various CD values. The second column shows SEM images of the corresponding printed patterns with CDs of 40 nm, 80 nm and 50 nm. The third column shows magnified views of the SEM image and the contours of the expected aerial images.

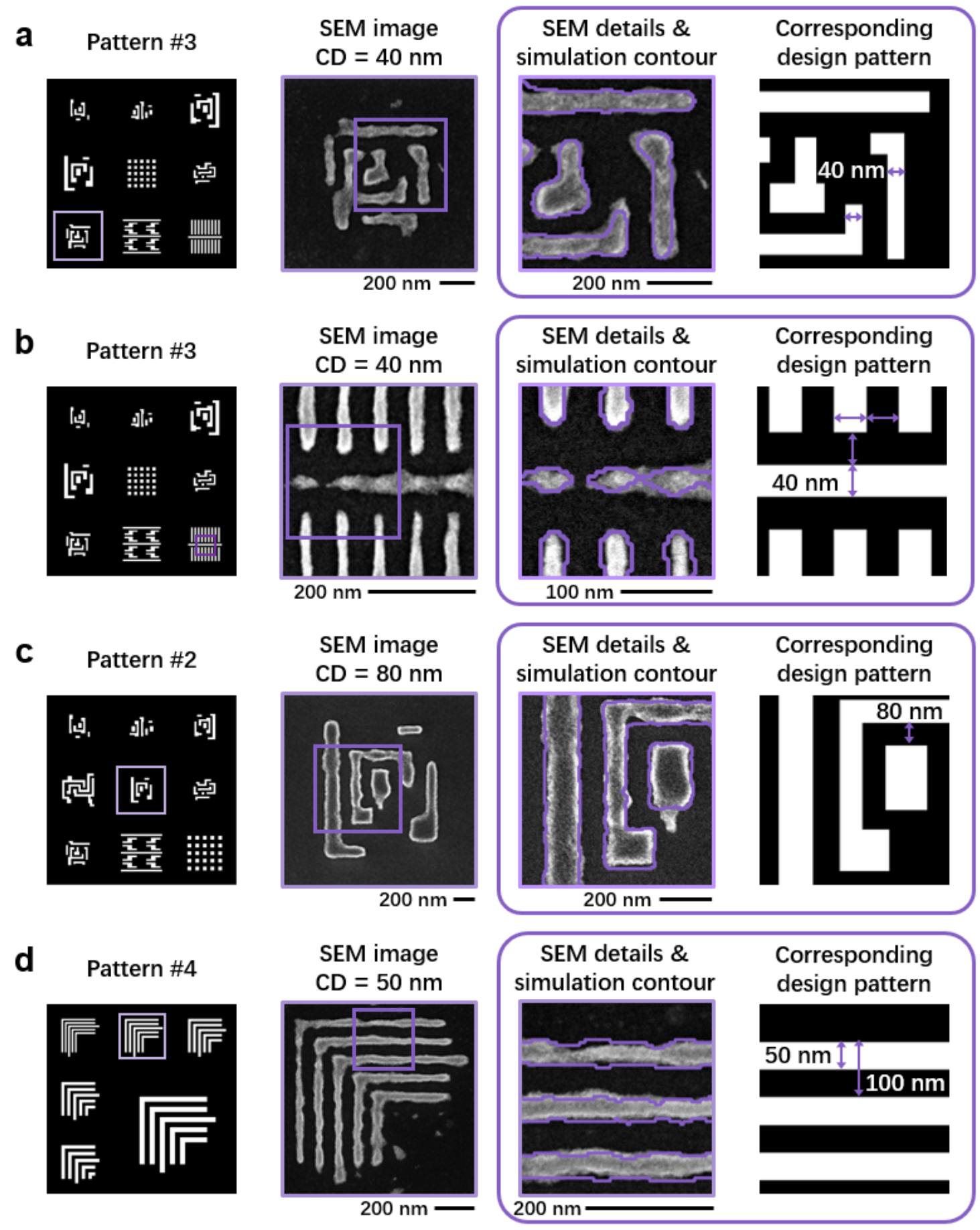

*Figure 3. (a) Glyph pattern with 40 nm CD, (b) Tip-to-line pattern with 80 nm pitch and 40 nm CD, (c) Glyph pattern with 80 nm CD and (d) elbow patterns with 50 nm CD and 100 nm pitch. The edges on the zoomed SEMs denote the simulated resist contour, and the plots on the right show the corresponding target pattern.*

## 4 Design and fabrication of the holographic mask

Unlike its visible-light counterpart, EUV holography employs masks with absorber thicknesses that exceed the illumination wavelength. Combined with the partial coherence and finite spectral bandwidth of realistic EUV sources, this regime demands an accurate physical model of the light–mask interaction during the design process. In this section we first discuss some basic optical constraints for the mask fabrication (Sec 4.1), then we will describe the imaging model used to compute the aerial image from a given mask layout (Sec. 4.2), we will quantify mask three-dimensional (M3D) effects and polychromatism whose effect can strongly degrade EUV-HL imaging if not properly modeled (Sec. 4.3), and finally we will present the computer-generated hologram (CGH) algorithm and the mask fabrication process (Sec. 4.4).

### 4.1 Holographic mask size and resolution

To evaluate the approximate resolution limit of a holographic mask, consider a target image consisting of an isolated dot with diameter $d$. In the idealized monochromatic paraxial case, the holographic mask that generates it is a Fresnel zone plate with outermost zone width $w = d/1.22$[28], which would suggest that the minimum feature size on the mask sets a hard floor on the achievable resolution. This identification of mask feature size with zone width, however, is specific to the zone-plate geometry. A photon sieve, in which the continuous Fresnel zones are replaced by discrete pinholes distributed over the same zone pattern, breaks this correspondence: the pinhole *diameter* is the smallest feature that needs to be fabricated, while their *center positions* still encode the underlying zone structure. Because pinholes can be placed on zones narrower than themselves, and because the focal contribution of a pinhole peaks for a diameter of roughly 1.5 times the local zone width[29], the resolution of a photon sieve is not bounded by its smallest fabricated feature[30]. Menon et al. demonstrated this experimentally in a lithographic context, exposing lines roughly half the diameter of the smallest pinhole in the sieve[31]. The holographic mask design approach described in this work generalizes this idea further: the mask layout is the result of an inverse-design optimization that is free to place absorber features anywhere on the membrane, with no requirement that the smallest fabricated feature coincide with the effective resolution element of the diffracted image. The relation $d = 1.22w$ should therefore be read as a useful order-of-magnitude estimate rather than a fundamental bound, and a suitable design can in principle extend the printed resolution below it.

We emphasize that the masks demonstrated in the present work do not yet exploit this margin: the smallest mask feature is 40 nm and the smallest printed CD is also 40 nm, so the experimental results sit essentially at the zone-plate limit. Pushing the printed CD below the minimum fabricated feature is a natural next step, and one that is fully compatible with our inverse-design framework, since the optimizer already operates on a pixel grid that need not be matched to the target resolution.

The diameter of a zone plate with focal length $f$ and resolution $d$ is $D = 2f \cdot \tan(\arcsin(\lambda/2w))$. To print a pattern over a given field of view $FOV$, the mask size $M$ must therefore satisfy $M = FOV + D$. This means that achieving finer resolution requires a proportionally larger mask, and that the mask size grows with working distance. In the paraxial regime ($\lambda \ll w$), this simplifies to $D \approx f\lambda/w$, directly analogous to the role of numerical aperture in projection lithography.

## 4.2 EUV-HL imaging model

Figure 5(a) sketches the imaging procedure. The polychromatic EUV beam illuminates the transmissive HL mask, and for each wavelength component a corresponding near field (NF) is generated at the object plane. Rigorous electromagnetic simulation of the full mask is prohibitively expensive at the mask sizes used in this work, so we adopt a convolution-based M3D model: the local 3D scattering response of the absorber stack is captured in a kernel obtained from a rigorous simulation of a small, representative region of the mask, and the NF over the full mask is then approximated by convolving this kernel with the mask pattern. This treats the mask as a shift-invariant system, an approximation that is justified when the M3D response decays over a length scale shorter than the characteristic variation of the mask pattern, as is the case here.

Each monochromatic NF then propagates in free space to the wafer plane, where the corresponding aerial-image component is computed using the angular-spectrum method. The total intensity distribution at the wafer is obtained as the incoherent summation of all monochromatic aerial-image components, weighted by the source spectrum. Finally, the resist response is modeled by a sigmoid function that represents the binarization of the latent image upon exposure and development, yielding the predicted resist pattern R.

To quantify the imaging quality of a given mask design, we used the MSE defined in section 3, but here T is the target Boolean layout, R is the simulated resist pattern computed as described above and N is the total number of the pixels in the array. The MSE is used both as a figure of merit when assessing the impact of physical effects in Sec. 4.3 and as the loss function in the CGH optimization in Sec. 4.4.

## 4.3 Critical physical effects

Two physical effects dominate the imaging fidelity of EUV-HL and must therefore be captured in both the simulation and the mask design: M3D effects and source polychromatism. We discuss each in turn, then briefly address how the limited focus accuracy of our exposure setup is handled. Less critical effects are treated in the Supplementary Material.

**M3D effects.** Under EUV illumination, the absorber thickness is comparable to or larger than the wavelength, so the mask topography strongly perturbs the transmitted near field. The thin-mask approximation, which assigns NF amplitudes of 1 and 0 to transparent and absorbing regions respectively, is inadequate in this regime. Figure 4(a) compares aerial images simulated for the same mask layout under the thin-mask approximation and under the full M3D convolution model: the thin-mask result deviates strongly from the target elbow pattern, while the M3D result reproduces it faithfully and matches the experimental SEM image. Figure 4(b) shows the corresponding MSE as a function of defocus for both models. The M3D-based simulation yields a low MSE at best focus that grows smoothly with defocus, as expected. The thin-mask simulation, by contrast, exhibits a high MSE across the full defocus range, confirming that neglecting M3D effects systematically corrupts the predicted aerial image regardless of focus position.

**Polychromatism.** Because EUV-HL relies on diffraction rather than imaging through purpose-designed optics, it is highly sensitive to the spectral bandwidth of the illumination. Figure 4(c) illustrates this. In the upper row, an HL mask optimized assuming monochromatic illumination produces an excellent aerial image when simulated at the design wavelength (MSE = 0.0028), but the image degrades severely when re-simulated using the actual ~4% bandwidth of the synchrotron source — the pattern becomes too blurred to resolve discernible features. In the lower row, an HL mask optimized directly against the polychromatic source spectrum yields an aerial image of satisfactory fidelity under realistic conditions (MSE = 0.0092). Two consequences of this are worth noting. First, the wavelength sampling used during optimization must be fine enough to resolve the spectral features of the source: as shown in Fig. 4(d), the MSE rises rapidly as the sampling becomes coarser and approaches the monochromatic limit. Second, because chromatic aberration predominantly degrades the high-spatial-frequency components of the aerial image, optimizing for a polychromatic spectrum implicitly suppresses sensitivity to high-frequency phase shifts.

**Focus accuracy of the exposure setup.** Defocus is not a fundamental limitation of EUV-HL: with a suitable optomechanical assembly, the wafer plane can be positioned with sub-nanometer precision, as routinely achieved in commercial EUV scanners. In the interference-lithography end station used in this work, however, the wafer plane can be set with high precision but with a comparatively low absolute accuracy of approximately 100 nm. To accommodate this setup-specific tolerance without compromising the printed patterns, we include the expected focus uncertainty as a parameter in the CGH loss function, so that the optimization yields a mask whose aerial image remains close to the target across the relevant defocus window. The MSE-versus-defocus curve in Fig. 4(b) shows the resulting tolerance: the aerial image of a representative HL pattern remains close to the target across a focus range that comfortably covers the ±100 nm uncertainty of our setup. Implementation details of the focus-tolerant loss function are given in the Supplementary Material.

Together, Figure 4 demonstrates that without proper modeling of M3D effects and source polychromatism, the simulated aerial image deviates substantially from the experimental result, and any mask designed against such a degraded model will fail to print the intended pattern. All scale bars in Fig. 4 are 200 nm.

### 4.4 CGH algorithm and mask fabrication

The HL mask layout is computed by inverse optimization against the imaging model of Sec. 4.2. The MSE between the predicted resist pattern and the target layout is used as the loss function, and the mask layout is updated iteratively by gradient descent.

Starting from a uniform-value initial guess and adding random white noise when the optimization tends to stagnate, the algorithm produces a continuous-tone mask that is binarized in the final step to satisfy the manufacturability constraint of a single-thickness absorber. A rigorous NF calculation, which would require a full 3D electromagnetic simulation (FDTD or RCWA) of the entire mask, is intractable at the scale of our holographic masks. We therefore approximate the M3D response as a shift-invariant operator: a convolution kernel is extracted from a small, representative portion of the training mask[32], which is simulated based on Born series, a fast numerical method for solving Helmholtz equation[33–35], then convolved with the full mask pattern to obtain the full-mask NF. The propagation from the mask to the wafer plane is then carried out exactly with the angular spectrum method, which is itself a shift-invariant operation. Because both steps reduce to FFT-based convolutions, the forward and adjoint computations required by the optimization run roughly $10^4$ times faster than an equivalent rigorous electromagnetic-field simulation of the full mask, at the cost of approximating the local M3D response. The loss function can also be customized to address specific design requirements. For example, jointly optimizing the imaging quality at multiple focal planes to enlarge the depth of focus, as used in this work. A detailed derivation of the gradients is given in the Supplementary Material.

The performance of EUV-HL depends critically on how faithfully the optimized mask layout is reproduced in the fabricated mask, since any deviation from the designed geometry, like line-edge roughness, CD bias, or sidewall slope, directly perturbs the diffracted field and therefore the aerial image. To minimize fabrication-induced nonidealities, we adopted a direct-patterning approach in which the HSQ layer serves simultaneously as the e-beam resist and as the EUV absorber, eliminating any subsequent pattern-transfer step. This strategy is inspired by the established fabrication of gratings for EUV interference lithography, where resist layers are patterned directly into line–space arrays without further processing. HSQ is well suited to this dual role: it is stable under EUV irradiation and supports high-resolution patterning by e-beam lithography. The resulting mask stack is shown schematically in Fig. 5(b) and consists of a 6 × 6 $mm^2$ Si support frame with a 2 × 2 $mm^2$ $Si_3N_4$ membrane of 80 nm thickness. The design is patterned at the center of the membrane using a 200 nm HSQ resist layer. The pixel size of 40 nm creates features with aspect ratio of 5:1, increasing the risk of pattern collapse due to aqueous capillary forces, therefore, necessitating the use of critical point drying (CPD) after resist development. SEM characterization of the masks (see Fig. 1(b)) includes both the pixel size and the shaping quality. The transmission of this HSQ layer at 13.5 nm is about 13.5% which is not negligible, therefore the complex index of refraction of the material must be accounted for during the rigorous propagation step.

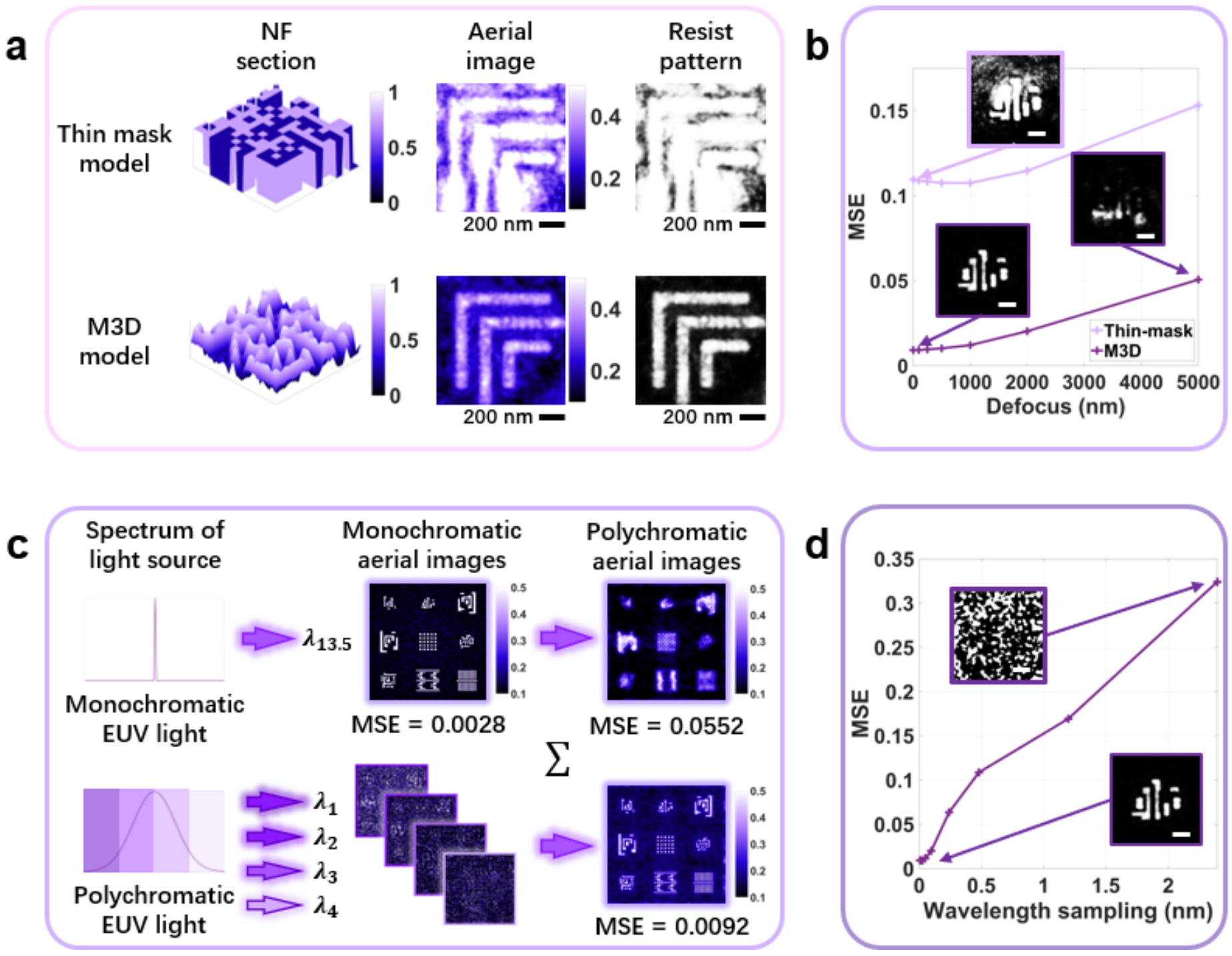

Figure 4. (a) The sketch and the impact of M3D effect. The mask topography strongly impacts the NF distribution, which degrades the simulation accuracy without appropriate modeling. (b) plot of aerial image MSE versus defocus under M3D model and thin mask simulation. As expecte4d, the aerial image MSE rises as the defocus increase. (c) Impact of chromatism on the image quality. The image intensity is calculated as the incoherent summation of monochromatic aerial images. By taking the chromatism into account during the design, EUV HL can generate appropriate aerial image under polychromatic illumination. (d) Plot of MSE as a function of the wavelength sampling. As the sampling becomes coarser the MSE grows rapidly.

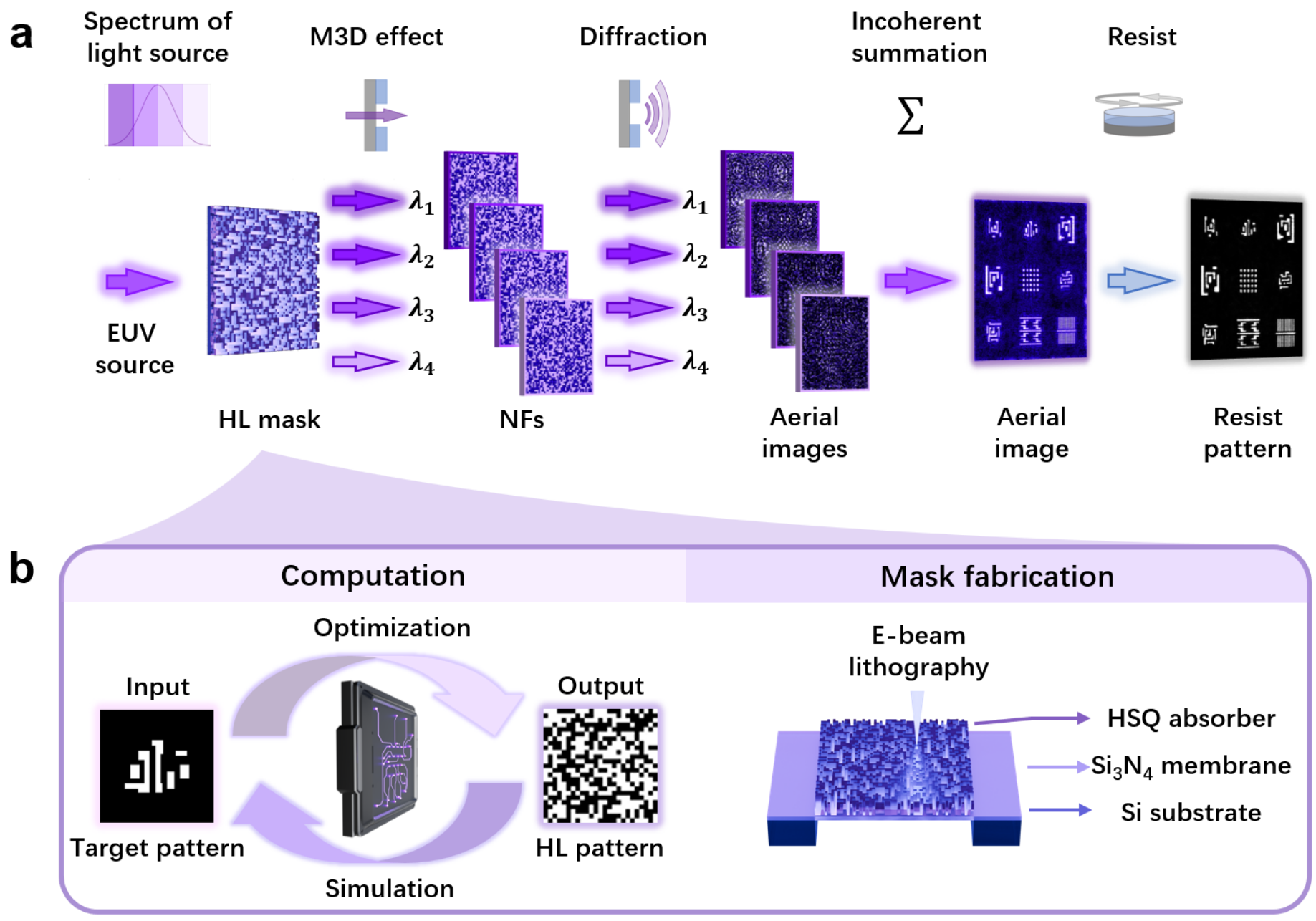

**Fig. 5.** (a) The process of EUV HL imaging model. Given the spectrum information of the illumination, monochromatic NFs, and their corresponding aerial images are calculated. The total intensity distribution is the incoherent summation of all monochromatic aerial images, and the resist pattern is finally calculated with a sigmoid function model. (b) EUV HL mask computation and fabrication: The calculation of an HL mask is an iterative optimization process; the transmissive EUV HL mask is then fabricated by e-beam lithography. The mask is composed of Si substrate, a 80 nm thick $Si_3N_4$ membrane, and a patterned 200 nm HSQ layer.

## 5 Conclusion and outlook

We have demonstrated EUV holographic lithography at 40 nm critical dimension on arbitrary, non-periodic, curvilinear layouts achieving nearly an order-of-magnitude improvement over the previous best EUV-HL resolution of 372 nm on periodic elbow patterns[24,26]. Three elements were essential to closing this gap: an imaging model that captures the dominant physical effects of EUV mask diffraction, in particular M3D scattering and source polychromatism; an inverse-design CGH algorithm that exploits the shift-invariance of the model to optimize full mask layouts efficiently; and a direct-write HSQ fabrication process that transfers the optimized layout to the mask without intermediate pattern-transfer steps.

The result is a lensless EUV patterning method that fills a design space gap that neither projection EUV nor EUV-IL can cover. Projection EUV provides resolution and throughput for high-volume manufacturing, but the cost and complexity of its reflective optics place it out of reach for academic and small-industrial research. EUV-IL provides a lensless, low-cost route to sub-10 nm features, but only for periodic geometries. EUV-HL, as demonstrated here, combines the lensless simplicity of EUV-IL with the arbitrary-pattern freedom of projection EUV, at a resolution sufficient for a broad class of device-relevant geometries.

This positions EUV-HL to address two needs that are not currently well served. First, arbitrary, non-periodic, curvilinear features at sub-50 nm scales are precisely what is required for the prototyping of metasurfaces, photonic crystals, single-photon emitter arrays, and superconducting nanowire structures, geometries that are increasingly central to photonic and quantum technologies and that interference lithography cannot provide. Second, because the holographic mask and the inverse-design framework are largely decoupled from the specific illumination wavelength, the method can be transferred to beyond-EUV (BEUV, ~3.x nm) wavelengths with comparatively modest optical changes. This opens a clear path to non-periodic BEUV pattern prototyping and resist screening, an early-stage capability that is currently inaccessible to BEUV-IL and that will be needed before BEUV process development can mature.

Several limitations remain. The transmissive membrane geometry used here will not scale straightforwardly to wafer-scale fields; large-area, mechanically robust membranes or a reconfiguration to reflective masks will be required for larger printed fields of view. Higher temporal coherence, for example from a free-electron laser source[28] and high-order harmonic generation source[36,37], would relax the polychromatic-design constraint and allow finer features. Resists with higher EUV sensitivity, particularly metal-oxide and inorganic platforms[38], would shorten exposure times and improve line-edge roughness. Multi-beam mask writers would substantially reduce the e-beam time required to fabricate large CGH masks. None of these constraints is fundamental to the method, and each maps onto an active line of development in the broader EUV community.

Taken together, this work establishes EUV holographic lithography as a sub-50 nm patterning technique with demonstrated performance on device-relevant geometries, and as the natural lensless complement to interference lithography for arbitrary EUV-scale design. It is, today, a uniquely flexible tool for nanostructure prototyping at 13.5 nm, and a clear enabler for the non-periodic BEUV pattern development that will be needed in the coming years.

## Author contributions

Ziqi Li conceived the idea and developed the computation algorithm; Iason Giannopoulos fabricated the HL mask and conducted the experiment; Iacopo Mochi supervised and supported the whole project; Iacopo Mochi, Iason Giannopoulos, Ziqi Li, Dimitrios Kazazis designed the experiment; Yayi Wei, Lisong Dong, Dimitrios Kazazis, Yasin Ekinci, and Xu Ma supervised part of the project; Zongqiang Yu and Zhiyuan Niu supported part of the experiment; Iacopo Mochi, Ziqi Li, and Iason Giannopoulos wrote the manuscript. All authors discussed the results and revised the manuscript.

## Data availability

All data are available from the corresponding authors upon reasonable request.

## Conflicts of interest

There are no conflicts of interest to declare.

## Acknowledgements

Part of this work was performed at the multi-functional beamline and endstation (BL05U-A1) of the National Synchrotron Radiation Laboratory in Hefei, China. The authors express special thanks to Zhao Wu, Xiaolei Wen, Aoqi Zhu for their support and to Joan Vila Comamala for his insightful input on diffractive optics.